\documentclass[print]{ieeecolor}
\usepackage{generic}
\usepackage{cite}
\usepackage{amsmath,amssymb,amsfonts}
\usepackage{algorithmic}
\usepackage{graphicx}
\usepackage{stmaryrd}
\usepackage{textcomp}
\usepackage{subfig}
\usepackage{hyperref}
\newcommand{\uproman}[1]{\uppercase\expandafter{\romannumeral#1}}

\begin{document}
	\title{CutFEM forward modeling for EEG source analysis}
\author{%
*\textsuperscript{1}Tim Erdbrügger, 
\textsuperscript{1,2}Andreas Westhoff, 
\textsuperscript{1}Malte Höltershinken,
\textsuperscript{3,4}Jan-Ole Radecke, 
\textsuperscript{1}Yvonne Buschermöhle, 
\textsuperscript{5}Alena Buyx,  
\textsuperscript{6}Fabrice Wallois, 
\textsuperscript{7}Sampsa Pursiainen,
\textsuperscript{1}Joachim Gross,
\textsuperscript{3,4}Rebekka Lencer, 
\textsuperscript{2}Christian Engwer,
\textsuperscript{1}Carsten Wolters%
\thanks{Acknowledgement: This work was supported by the Bundesministerium für Gesundheit (BMG) as project ZMI1-2521FSB006, under the frame of ERA PerMed as project ERAPERMED2020-227 and by the Deutsche Forschungsgemeinschaft (DFG), projects WO1425/10-1, GR2024/8-1, LE1122/7-1.
C. Engwer was supported by the Deutsche Forschungsgemeinschaft (DFG, German Research Foundation) under Germany’s Excellence Strategy EXC 2044-390685587, Mathematics Münster: Dynamics–Geometry–Structure.}
\thanks{\textsuperscript{1}Institute for Biomagnetism and Biosignalanalysis, University of Münster, Malmedyweg 15, 48149, Münster (email: tim.erdbruegger@uni-muenster.de).
}
\thanks{\textsuperscript{2}Institute for Analysis and Numerics, University of Münster, Germany.}%
\thanks{\textsuperscript{3}Dept. of Psychiatry and Psychotherapy, University of Lübeck, Ratzeburger Allee 160, 23562, Lübeck, Germany.}
\thanks{\textsuperscript{4}Center for Brain, Behaviour and Metabolism (CBBM), University of Lübeck, Ratzeburger Allee 160, 23562, Lübeck, Germany.}%
\thanks{\textsuperscript{5}Institute of History and Ethics in Medicine, Technical
University of Munich, Germany.}%
\thanks{\textsuperscript{6}Institut National de la Sant\'{e} et de la Recherche M\'{e}dicale, University of Picardie Jules Verne, France.}
\thanks{\textsuperscript{7}Computing Sciences Unit, Faculty of Information Technology and Communication Sciences, Tampere University, Finland.}
}
	\maketitle
	\begin{abstract}
    Source analysis of Electroencephalography (EEG) data requires the computation of the scalp potential induced by current sources in the brain. This so-called EEG forward problem is based on an accurate estimation of the volume conduction effects in the human head, represented by a partial differential equation which can be solved using the finite element method (FEM). 
    FEM offers flexibility when modeling anisotropic tissue conductivities but requires a volumetric discretization, a mesh, of the head domain.
    Structured hexahedral meshes are easy to create in an automatic fashion, while tetrahedral meshes are better suited to model curved geometries. Tetrahedral meshes thus offer better accuracy, but are more difficult to create.
	
	\textit{Methods:} 
    We introduce
    CutFEM for EEG forward simulations
    to integrate the strengths of hexahedra and tetrahedra.
    It belongs to the family of unfitted finite element methods, decoupling mesh and geometry 
    representation.
    Following a description of the method, we will employ CutFEM in both controlled spherical scenarios and the reconstruction of somatosensory evoked potentials.
		
	\textit{Results:} CutFEM outperforms competing FEM approaches with regard to numerical accuracy, memory consumption and computational speed while being able to mesh arbitrarily touching compartments.
		
	\textit{Conclusion:} 
    CutFEM balances numerical accuracy, computational efficiency and a smooth approximation of complex geometries that has previously not been available in FEM-based EEG forward modeling.
		
	\end{abstract}
    
    \section{Introduction}
	Electroencephalography (EEG) is a widely used tool for the assessment of neural activity in the human brain \cite{Brette2012}. To estimate the area of the brain responsible for the measured data, one has to simulate the electric potential as induced by hypothetical current sources in the brain, i.e. the EEG forward problem has to be solved. While quasi-analytical solutions to the differential equation underlying the forward problem exist, these are only available in simplified geometries such as the multi-layer sphere model\cite{DeMunck1993}. One thus requires numerical methods 
 to incorporate 
 accurate representations of the head's shape and volume conduction properties. 
 Popular 
 approaches are the boundary element method (BEM)\cite{Mosher1999,Gramfort2011,Makarov2020} and the finite element method (FEM)\cite{Medani2015,Vallaghe2010}. Here, we will focus on the FEM due to its flexibility in modeling complex geometries with inhomogeneous and anisotropic compartments \cite{He2020,Vermaas2020,Beltrachini2018a,VanUitert2004,Schimpf2002,Wolters2007,Nuesing2016}.
    Efficient solvers and the transfer matrix approach \cite{Lew2009,Wolters2004}
    allow significantly reduced computational costs.
	
	When 
 employing FEM, 
 one usually choses between either a hexahedral or tetrahedral discretization of the head. Both choices come with their own strengths and limitations. The mesh creation requires a classification of the MRI into tissue types. This segmentation data often comes in the form of binary maps with voxels of around 1mm resolution, allowing for quick and simple hexahedral mesh generation. However, as head tissue surfaces are smooth, approximating them with regular hexahedra is bound to be inaccurate. While methods for geometry adaptation exist \cite{Wolters2007}, the resulting meshes still have a (reduced) angular pattern. Furthermore, when applying a standard continuous Galerkin FE scheme, areas with very thin compartments may suffer from leakage effects where current can bypass the insulating effects of the skull \cite{Sonntag2013}. To alleviate this, flux based methods, like the discontinuous Galerkin method, offer a robust alternative \cite{Engwer2017a}. These however severely increase the number of DOF and thus necessary computational effort.
	
	Surface-based tetrahedral FEM approaches on the other hand are able to accurately model the curvature of smooth tissue surfaces. Creating high quality tetrahedra, e.g. ones fulfilling a delaunay criterion, require tissue surface representations in the form of triangulations first. These triangulations have to be free of self-intersections and are often nested, usually leading to modeling inaccuracies such neglecting skull holes or an artificial separation of gray matter and skull. Therefore, we will not discuss surface-based tetrahedral FEM approaches throughout this work. 

    In \cite{Rice2013}, the impact of prone vs supine subject positioning on EEG amplitudes was investigated. In the small group study, average differences of up to eighty percent were found. These were accompanied by differences in MRI-based CSF-thickness estimation of up to thirty percent underlining the importance of correctly modeling CSF-thickness and areas of contact between skull and brain surfaces.
	
	Recently, an unfitted discontinuous Galerkin method (UDG) \cite{Bastian2009} was introduced to solve the EEG forward problem\cite{Nuesing2016}. Rather than working with mesh elements that are tailored to the geometry, it uses a background mesh which is cut by level set functions, each representing a tissue surface. It was shown to outperform the accuracy of a discontinuous Galerkin approach on a hexahedral mesh, while not being limited by the assumptions necessary to create tetrahedral meshes.

	Extending the ideas of the UDG method, this paper introduces a multi-compartment formulation of the CutFEM \cite{Burman2015} for EEG source analysis. Compared to UDG, it operates on a simpler trial function space and adds a ghost penalty based on \cite{Burman2010}, stabilizing small mesh elements. 
 
    The paper is structured as follows. After introducing the theory behind CutFEM, three successively more realistic scenarios are tested. These scenarios include a multi-layer sphere model, followed by realistic brain tissues embedded in spherical skull and scalp compartments. Finally, a fully realistic five compartment head model is used for source analysis of the P20/N20 component of measured somatosensory evoked potentials (SEP). Comparison results from different FEM and meshing approaches will be considered throughout the scenarios.
	
	\section{Theory}
	\subsection{CutFEM}
	Deviating from classical, fitted FEM-approaches, where 
 the mesh cells resolves tissue boundaries, 
 CutFEM uses a level set based representation of domain surfaces. 
	Let $\Omega = \bigcup_i \Omega_i$ be the head domain divided into $m$ disjunct open subdomains, e.g. gray matter, white matter, CSF, skull, skin. The level set function for compartment $i$ is then defined as
	$$ \Phi_i(x) \left\{ \begin{array}{ l l }
	< 0, \text{ if $x \in \Omega_i$} \\
	= 0, \text{ if $x \in \partial \Omega_i$} \\
	> 0, \text{ else}
	\end{array} \right.
	$$
	and $\mathcal{L}_i = \{x \in \Omega : \Phi_i(x) = 0\}$ 
 denotes its (zero) level set. 
	We proceed by defining a background domain $\hat{\Omega} \subset \mathbb{R}^3$ covering the head domain $\Omega$. This background is then tesselated, yielding a regular hexahedral mesh $\mathcal{T}(\hat{\Omega})$, the fundamental or background mesh.
	Taking on the level set representation, submeshes $\mathcal{T}_h^i \subset \mathcal{T}_h(\hat{\Omega}) $ are created from the background mesh, containing all cells that have at least partial support within the respective subdomain $\Omega_i$. This results in an overlap of submeshes at compartment interfaces. For each submesh we define a conforming $\mathbb{Q}_1$ space $V_h^i$. Thus, up to this point, each submesh is treated the way a conforming Galerkin method would treat the entire mesh.
	
	The difference then lies in restricting the trial and test functions to their respective compartment, effectively cutting them off at the boundary and giving rise to the name CutFEM.
    A fundamental mesh cell intersected by a level set $\mathcal{L}_i$ is called a cut cell. Their respective fundamental cells are contained in multiple compartements and thus have more DOF.
    On the other hand, compared to classical conforming discretizations, a coarser mesh resolution can be chosen, as the mesh does not have to follow small geometric features.

	As the trial functions are only continuous on their respective compartment and cut off at the boundary, using them to approximate the electric potential requires internal coupling conditions at the tissue interfaces.
	We 
 define the internal skeleton as the union of all subdomain interfaces
	\begin{equation}
    \begin{split}	    
	\Gamma = \bigcup \Big\{\bar\Omega_i \cap \bar\Omega_j :
 ~i \neq j,~\mu_{d-1}(\bar\Omega_i \cap \bar\Omega_j)>0 \Big\}.\label{eq:1}
	\end{split}
 \end{equation}
	$\mu_{d-1}$ is the d-1 dimensional measure in d-dimensional space. For two sets $E,F$ sharing a common interface (an element of $\Gamma$) and a, possibly discontinuous, function $u$ operating on them we can define a scalar- or vector-valued jump operator as $\llbracket u \rrbracket := u|_E \cdot n_E + u|_F \cdot n_F$ with $n_E, n_F$ the outer unit normal of the respective set. Additionally, a (skew-)weighted average can be stated as
	\begin{align}
	\{u  \} &= \omega_E u|_E + \omega_F u|_F\label{eq:2}\\
	\{u  \}^* &= \omega_F u|_E + \omega_E u|_F.\label{eq:3}
	\end{align}
	with 
	$\omega_E = \frac{\delta_E}{\delta_E + \delta_F}$,
	$\delta_E = n_E^t\sigma_E n_E.$
	Here, $\sigma_E$ refers to the symmetric $3 \times 3$, positive definite electric conductivity tensor on $E$. Note that $	\llbracket uv \rrbracket  = \llbracket u\rrbracket \{v \}^* + \{u \} \llbracket v \rrbracket.$ The purpose of these definitions will become clear when deriving the weak formulation for our forward model.
	
	Typically, the EEG forward problem for the electric potential $u$ induced by a neural source term $f$ is derived from the quasi-static formulation of Maxwell's equations  \cite{Brette2012}.
	\begin{align}
	\nabla \cdot \sigma \nabla u &= f, \; \text{ in } \bigcup\limits_i \Omega_i\label{eq:4}\\
    \langle \sigma \nabla u,n\rangle &= 0, \; \text{ on } \partial\bar{\Omega}\label{eq:5}\\
    \intertext{And in addition we require continuity of the electric potential and the electrix current}
    \llbracket u \rrbracket &= 0, \; \text{ on } \Gamma\label{eq:6}\\
	\llbracket \sigma \nabla u \rrbracket &= 0, \; \text{ on } \Gamma. \label{eq:7}
	\end{align} 
	As trial and test space we employ $V_h$ as direct sum of all $V_h^i$. 
	
	The weak formulation can be obtained by multiplying with a test function, integrating and applying subdomain wise integration by parts. This yields
	$$
	\sum_{i}(\int_{ \Omega_i} \sigma \nabla u_h^i \nabla v_h^i - f v_h^idx)  -  \int_{\Gamma} \{\sigma \nabla u_h \}\llbracket v_h\rrbracket dS = 0,
	$$
	where the jump formula for a product of two functions as well as \eqref{eq:7} were used.
	$u_h^i$ is the restriction of $u_h \in V$ to $V_i$. A symmetry term $ \pm  \int_{\Gamma} \{\sigma \nabla v_h \}\llbracket u_h\rrbracket dS$ is added to end up with either a symmetric or non-symmetric bilinearform.
	
	To incorporate \eqref{eq:6} 
	a Nitsche penalty term\cite{Nitsche1971} is added, which weakly couples the domains and ensures coercivity\cite{Burman2015}:
	\begin{equation}
	P_{\gamma}(u,v) = \gamma \nu_k\int_{\Gamma} \frac{\hat{\sigma}}{\hat{h}} \llbracket u_h\rrbracket \llbracket v_h\rrbracket dS.\label{eq:8}
	\end{equation}
    Here
    $\nu_k, \hat{h}, \hat{\sigma}$ are scaling parameters based on the ratio of cut cell area on each interfaces' side, dimension, degree of trial functions used and conductivity. See \cite{DiPietro2012} for a further discussion of these. $\gamma$ is a free parameter to be discussed later. 
	
	A challenge is the shape of the cut-cells. Heavily distorted and sliver-like snippets significantly impact the stability
    and
    lead to a deterioration of the stiffness matrix condition number. To alleviate this, a ghost penalty \cite{Burman2010} term is used, which weakly couples the solution on the snippets to their neighbors. Said coupling takes place on the interfaces of all the fundamental mesh cells cut by a level set. Let
    \begin{equation}
    \begin{split}
	\hat{\Gamma} = \cup \Big\{\partial E_i: E_i \in \mathcal{T}_h, E_i \cap \Gamma \neq \emptyset \}.\label{eq:9}
    \end{split}
	\end{equation}
	Note the difference between $\Gamma$ and $\hat{\Gamma}$. $\Gamma$  operates on compartment interfaces, $\hat{\Gamma}$ on faces of the fundamental mesh.
	The ghost penalty is then defined as
	\begin{equation}
	a^G (u_h,v_h) = \gamma_G \int_{\hat{\Gamma}} \hat{h} \llbracket \sigma \nabla u_h \rrbracket \llbracket \nabla v_h \rrbracket dS,\label{eq:10}
	\end{equation}
	where
	$\gamma_G$ is again a free parameter, usually a couple orders of magnitude smaller than $\gamma$. 
	
	The weak CutFEM EEG-forward problem can now be stated as finding the electric potential $u_h \in V_h$ such that
	\begin{equation}
	a(u_h,v_h) + a_{n/s}^N(u_h,v_h) + a^G(u_h,v_h) = l(v_h) \; \forall v_h \in V_h,\label{eq:11}
	\end{equation}
	with
	\begin{align*}
        a(u_h,v_h) &= \sum_{i} \int_{ \Omega_i} \sigma \nabla u_h^i \nabla v_h^idx,\\
        l(v_h) &=  \sum_{i} \int_{ \Omega_i} f v_h^idx
    \end{align*}
	and
	\begin{align*}
	a_{n/s}^{N}(u_h,v_h) :=  &- \int_{\Gamma}\{\sigma \nabla u_h \}\llbracket v_h\rrbracket  \pm \int_{\Gamma} \{\sigma \nabla v_h \} \llbracket u_h \rrbracket dS\\
	   &+ \gamma \nu_k\int_{\Gamma} \frac{\hat{\sigma}}{\hat{h}} \llbracket u_h\rrbracket \llbracket v_h\rrbracket dS.
	\end{align*}
	In the following we will refer to these two variants as NWIPG/SWIPG,
 short for non-symmetric/symmetric weighted interior penalty Galerkin method.
	
	In \cite{Oden1998,Guzman2009} it was shown that the non-symmetric DG-methods may result in a sub-optimal convergence rate in the L2-norm (full convergence in H1), a result that also extends to CutFEM\cite{Burman2012}.
	However, while SWIPG is coercive only if $\gamma$ is chosen sufficiently large\cite{Burman2012}, NWIPG does not have such a limitation.
	Therefore, we will employ the NWIPG method throughout this paper due to its stability with regard to the selection of $\gamma$.
 
	\paragraph{Integration over the cut domains}	
	Fundamental cells that are cut by level sets, the cut cells, can be integrated over by employing a topology preserving marching cubes algorithm (TPMC) \cite{Engwer2017}. The initial cell is divided into a set of snippets, each completely contained within one subdomain. These snippets are of a simple geometry and therefore easy to integrate over.  Thus, integrals over the fundamental cell or subdomain boundaries are replaced by integrals over the snippets or their boundaries. The trial functions are effectively cut off at the compartment boundaries.
	
	See Fig. 1 for an overview of the reconstruction steps. Note that the trial functions are coupled to their respective submesh, not to the TPMC reconstruction of the domain. The latter only determines the area over which the functions are integrated.
	
	\begin{figure}[t!]
		\centering
		\subfloat{\includegraphics[width=1\linewidth]{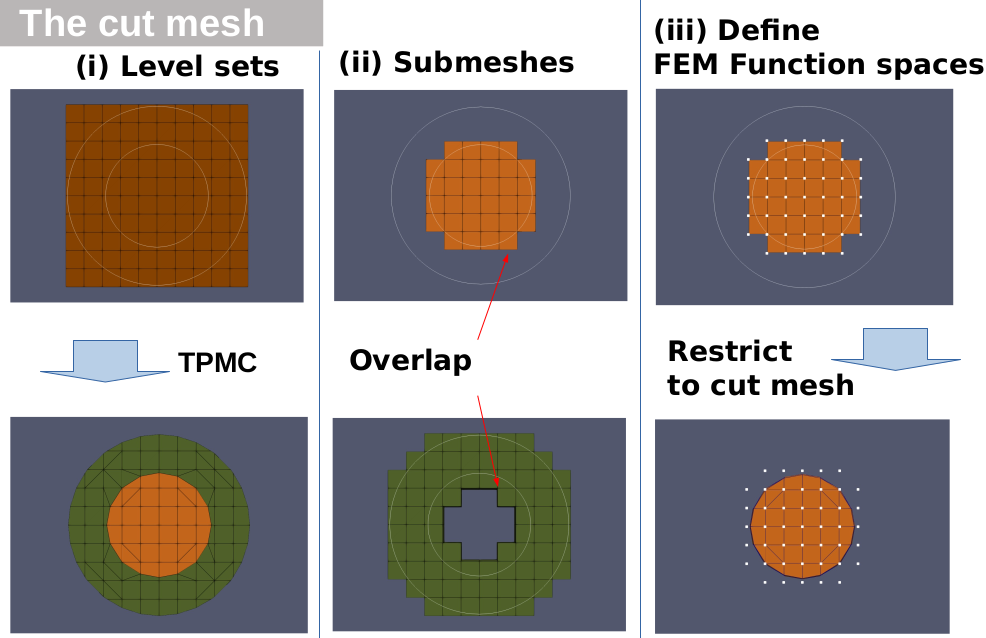}}
		\caption[Level sets over fundamental mesh and TPMC reconstruction]{Left: Fundamental mesh with two spherical level sets, topology preserving marching cubes reconstruction. Center: Overlapping submeshes for the two compartments enclosed by the level sets. Right: trial function space for the inner compartment with white dots representing degrees of Freedom, cut area that the DOF are restricted to.}
	\end{figure}
	
	Starting on the fundamental mesh, the algorithm is applied once per level set. Each following iteration is applied on the cut cells of the previous iteration, i.e. first the fundamental mesh is cut, then the resulting snippets are cut.  This ensures the correct handling of mesh cells that are cut by multiple level sets.

	\paragraph{Source model and transfer matrix}
	Following the principle of St. Venant, the source term $f$ will be approximated by a set of monopoles. Where fitted FEM use mesh vertices as monopole locations, this is not feasible for CutFEM as fundamental cells may have vertices not belonging to the source compartment. Rather, only gray matter cut cells are used and the locations are based on a Gauss-Legendre quadrature rule. For more information on the Venant source model, see \cite{Medani2015}, \cite{Buchner1997}.
	
	For an accurate source analysis it is necessary to compute the EEG-forward solution for a large number, i.e. tens of thousands, of possible sources. However, the electric potential induced by a source is only of interest at a set of predetermined points, namely the electrodes at the scalp. So, rather than solving \eqref{eq:11} for each source individually, a transfer matrix approach \cite{Gencer2004,Wolters2004} is employed, significantly reducing the amount of computation time needed.
	
	\section{Methods}
	

	\subsection{Head models}
	For numerical evaluations three progressively more realistic scenarios were created, two sphere models, one of which contains realistic brain tissues, and a five compartment model created from anatomical data. For each model, we will compare CutFEM and a geometry-adapted hexahedral CG-FEM approach (\textit{Hex}) with a node shift for the geometry-adaptation of 0.33 \cite{Wolters2007}. In the first model, the UDG approach of \cite{Nuesing2016} will also be added to the comparisons. To balance computational load, \textit{Hex} will use 1mm meshes whereas for CutFEM and UDG we use a 2mm background mesh.
	\paragraph{Shifted spheres}
	The first scenario contains the four spherical compartments brain, CSF, skull and scalp. The brain sphere will be shifted to one side, simulating a situation where the subject lies down and the brain sinks to the back of the skull. Conductivities were chosen according to \cite{McCann2019} with the exception that CSF and brain use the same conductivity. In terms of volume conduction the model is thus indistinguishable from a 3-layer concentric sphere model and analytical solutions \cite{DeMunck1993} are available as benchmark. Conductivity values and radii of the compartments can be found in Table 1.
	\begin{table}[htp]
		\begin{minipage}[c]{0.5\textwidth}
			\centering
			\begin{tabular}{l|ccc}
				& Radius [mm] & Center [mm]       & $\sigma$ [S/m] \\
                \hline
				Scalp & 92     & (127 127 127) & 0.43     \\
				Skull & 86     & (127 127 127) & 0.01     \\
				CSF   & 80     & (127 127 127) & 0.33     \\
				Brain & 78     & (129 127 127) & 0.33    \\[0.5ex]
			\end{tabular}
		\end{minipage}\hfill
		\begin{minipage}[c]{0.5\textwidth}
			\centering
			\caption[Radii, Center and Conductivities for the Shifted Sphere Model]{Radii, center and conductivities for the shifted sphere model.}
		\end{minipage}
	\end{table}
	
	 TPMC was applied twice, once on the fundamental mesh and once on the resulting cut cells. Note that this additional refinement step does not change the number of trial functions of the model. A total of 13.000 Evaluation points were distributed evenly throughout the inner sphere and lead fields for both radial and tangential source directions were computed at each point. For CutFEM, a combination of $\gamma = 16$ and $\gamma_G = 0.1$ has shown promising results. For UDG, no ghost penalty was implemented and $\gamma = 4$ was chosen, following {Nuesing2016}.
	
	\paragraph{Spheres containing realistic brain}
	In the previous section, the level set functions could be computed analytically up to an arbitrary accuracy. In a realistic scenario where the segmentation quality is limited by the MRI resolution as well as partial volume effects and MRI artefacts, this is not the case. An easy way to pass level sets to CutFEM lies in using tissue probability maps (TPM), a typical intermediate result \cite{Ashburner2014} from segmentation which provides for each voxel the probability that it is located in a certain compartment. 
	
	To examine the performance of CutFEM when used together with TPM's, another sphere model is employed, this time containing realistic gray and white matter compartments obtained from MRI scans of a human brain.	The subject was a healthy 24 year-old male from whom T1- and T2-weighted MRI scans were acquired using a 3 Tesla MR Scanner (MagnetomTrio, Siemens, Munich, Germany) with a 32-channel head coil. For the T1, a fast gradient-echo pulse sequence (TFE) using water selective excitation to avoid fat shift (TR/TE/FW = 2300/3.51 ms/8°, inversion pre-pulse with TI = 1.1 s, cubic voxels of 1 mm edge length) was used. For the	T2, a turbo spin echo pulse sequence (TR/TE/FA = 3200/408 ms/90°, cubic voxels, 1 mm edge length) was used. TPM's were extracted from both T1- and T2-MRI using SPM12 \cite{Ashburner2014} as integrated into fieldtrip \cite{Oostenveld2011}. For each voxel, the average of both TPM's was computed and a threshold probability of 0.4 was set as zero-line.
	
	The inner skull surface was defined as the minimal sphere containing the entire segmented brain with CSF filling the gaps. The spherical skull and scalp were chosen to have a thickness of 6mm. The same conductivities as before were used with CSF, gray and white matter being identical and again 200 sensors were placed on the scalp surface.
	\paragraph{Realistic 5 compartment head model}
	As an extension of the previous model, realistic 5 compartment head models were created using the same anatomical data, replacing the spherical skin, skull and CSF by realistic segmentations. Again, level sets were created from probability maps. To obtain smooth skull and scalp surfaces in the TPM case, binary maps of skull and skin were created following the procedure in \cite{Antonakakis2020}. The level sets of skull/skin were then calculated as an average of the binary map and the T1/T2 TPM again with a threshold of 0.4. Following \cite{Antonakakis2020}, the level sets were cut off below the neck to reduce computational load while maintaining a realistic current flow below the skull.
	Again, lead fields from hexahedral meshes were created for comparison. DOF, number of cut cells/mesh elements and the resulting number of snippets can be found in Table 2.
	\begin{table}[t!]
		\centering
		\begin{tabular}{l|lll}
			& DOF & Cut cells/elements & snippets         \\ \hline
			CutFEM & 917,463     & 716,994 & 7,950,120     \\
			$\textit{Hex}$ & 3,909,303     & 3,475,138 & -   
		\end{tabular}
		\centering
		\caption[Number of DOF, Cut cells and snippets]{Number of degrees of freedom/snippets/cut cells for CutFEM, number of degrees of freedom/elements for hexahedral mesh.}
	\end{table}
	\subsection{Forward and inverse comparisons}
	For the two spherical scenarios, analytical forward solutions were calculated as reference. For the realistic cases, somatosensory evoked potentials were recorded and a dipole scan was performed as described in detail in (b). 
	
	The two latter scenarios including realistic gray/white matter use a regular 2mm source grid created using Simbio \footnote{www.mrt.uni-jena.de/simbio}. It was ensured that the sources are located inside the gray matter compartment for both approaches (\textit{Hex} + CutFEM). The resulting source space contains 58.542 different dipole locations with no orientation constraint being applied.
	\paragraph{Error measures}
	
	Two different metrics were employed to quantify the observed errors, the relative difference measure (RDM) and the magnitude error (MAG) \cite{Wolters2007}. 
	
	The RDM measures the difference in potential distribution at the scalp electrodes.
	\begin{equation}
	RDM (\%)(u^{ana}, u^{num}) = 50*||\frac{u^{ana}}{||u^{ana}||_2} -\frac{u^{num}}{||u^{num}||_2}||_2.\label{eq:12}
	\end{equation}
	It ranges from $0$ to $100$, the optimal value being $0$.
	MAG determines the differences in signal strength at the electrodes.
	\begin{equation}
	MAG(u^{ana}, u^{num}) = 100*(\frac{||u^{num}||_2}{||u^{ana}||_2}-1).\label{eq:13}
	\end{equation}
	
	Measured in percent, its optimal value is $0$. It is unbounded from above and bound by $-100$ from below. $u^{ana}, u^{num} \in \mathbb{R}^s$ contain the analytical and numerical potential at the $s$ different sensor locations.
	
	CutFEM is implemented into the DUNEuro toolbox\footnote{https://www.medizin.uni-muenster.de/duneuro} \cite{Schrader2021} where the FEM calculations were performed. Analytical EEG solutions were calculated using the fieldtrip toolbox \cite{Oostenveld2011}.
	
	For a comparison of runtime and memory usage, the forward calculation is split into 5 steps. The time necessary to create a driver, i.e. the time DUNEuro needs to setup the volume conductor, the times needed to assemble the stiffness matrix and AMG solver, the transfer matrix solving process using Dune-ISTL \cite{Bastian2021} and the calculation of the final lead field matrix. All Computations are performed on a bluechip workstation with an AMD Ryzen Threadripper 3960X and 128 GB RAM. 16 threads are used to calculate the 200 transfer matrix/lead field columns in parallel.
	
	\paragraph{Somatosensory data and dipole scan}
	To investigate CutFEM's influence on source reconstruction, an electric stimulation of the median nerve was performed on the same subject the anatomical data was acquired from. 
	The stimuli were monophasic square-wave pulses of 0.5ms width in random intervals between 350-450ms. The stimulus strength was adjusted such that the right thumb moved clearly.
	EEG data was measured using an 80 channel cap (EASYCAP GmbH, Herrsching, Germany, 74 channel EEG plus additional 6 channels EOG to detect eye artifacts). EEG positions were digitized using a Polhemus device (FASTRAK, Polhemus Incorporated, Colchester, Vermont, U.S.A.).
	2200 stimuli were digitally filtered between 20 to 250Hz (50Hz notch) and averaged to improve signal to noise ratio. A single dipole scan was conducted over the whole source space using the data at the peak and the CutFEM lead field.
	
	The P20/N20 component typically exerts a high signal to noise ratio and a strongly dipolar topography, making it an ideal candidate for a dipole scan approach as motivated for example by \cite{Buchner1994}. 
	
    \begin{figure*}[tp]
		\centering
		\subfloat{\includegraphics[width=1\linewidth]{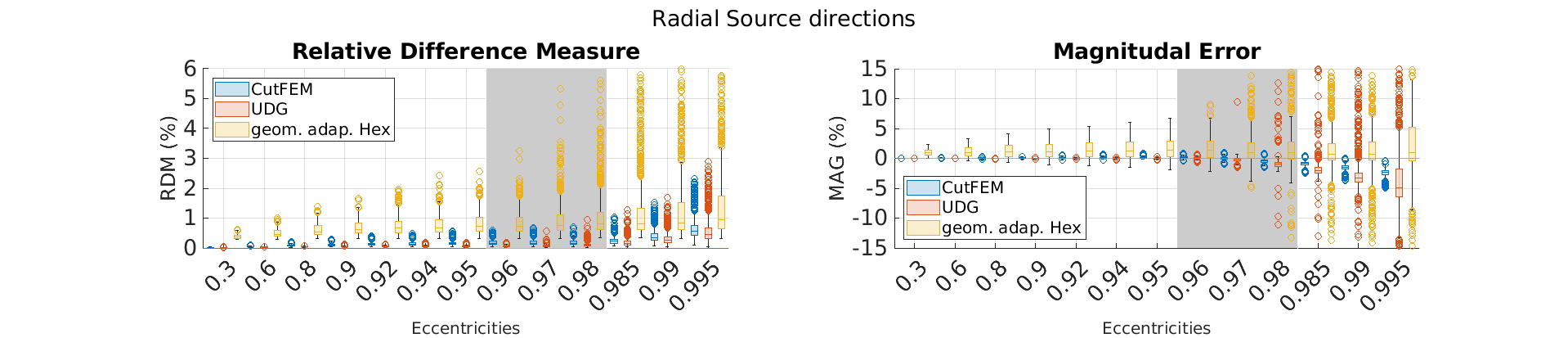}}
		\hfill
		\subfloat{\includegraphics[width=1\linewidth]{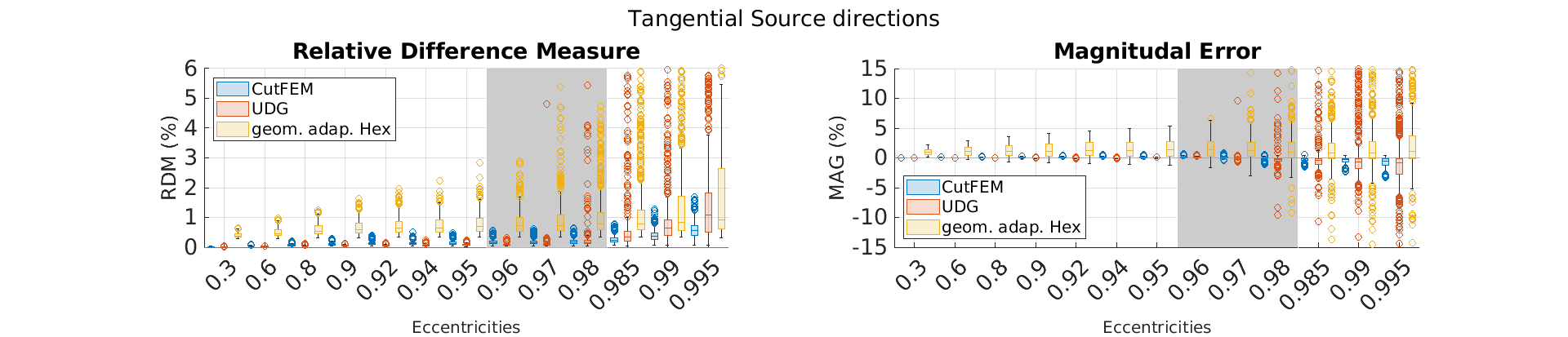}}
		\caption[EEG forward modeling errors for \textit{Hex} and unfitted FEM approaches]{EEG forward modeling errors for \textit{Hex} and unfitted FEM approaches in a shifted sphere scenario Top: Errors for tangential source directions. Bottom: Errors for radial source directions. Errors are in percent and grouped by eccentricities. The green line marks optimal error values. The grey area indicates the physiologically most realistic eccentricities.}
	\end{figure*}
	
\section{results}

    The first investigated model is the shifted sphere scenario, where the brain sphere was moved within the CSF-sphere until there was exactly one contact point between skull and brain (see \uproman{3} A). When comparing number of DOF and RAM usage, it is clear that CutFEM is by far the most memory efficient approach, using about a fifth of the number of trial functions and about a tenth of the amount of RAM as UDG (Table 3). \textit{Hex} also uses significantly more resources than CutFEM.
    
    Regarding computation time, as UDG has to solve a significantly larger system, each iteration step in the solution phase takes longer than for CutFEM. As most time is spent on solving the system, CutFEM is overall around 16 minutes or 34 percent faster than UDG. The same cannot be said for comparisons to the standard \textit{Hex} approach. While each iteration of the solver required less time than for \textit{Hex}, it required an average of 92 iterations compared to 14 for \textit{Hex}. The unfitted approaches spend less time calculating the final lead field as the time needed to locate each dipole within the 2mm background mesh is lower than for the 1mm hexahedral mesh. In total, the hexahedral CG was only faster than CutFEM by a negligible 3 percent or 52 seconds.
    
    \begin{table}[ht]
		\begin{minipage}[c]{0.5\textwidth}
			\centering
			\begin{tabular}{l|lll}
				               & CutFEM & UDG      & \textit{Hex}        \\ \hline
				number DOF    & 552 985   & 3 601 824 &  3 341 280 \\
				max. RAM used   & 6.91 GB   & 64.77 GB & 40.2GB \\ \hline
				Driver setup    & 44s       & 45s      & 52s     \\
				Matrix assembly & 319s      & 161s     & 25s     \\
				Solver setup    & 353s      & 235s     & 45s     \\
				Solving         & 1111s     & 2367s    & 1550s   \\
				Lead field      & 22s       & 20s      & 125s   \\ \hline
				Total time      & 1849s     & 2828s    & 1797s   \\

			\end{tabular}
		\end{minipage}\hfill
		\begin{minipage}[c]{0.4\textwidth}
			\centering
			\caption[Computation times, RAM/DoF usage]{Computation times, RAM/degree of freedom usage in the shifted sphere model. }
		\end{minipage}
	\end{table}
    
Error comparisons between CutFEM, UDG and \textit{Hex} can be found in Fig. 2. CutFEM outperforms \textit{Hex} in all eccentricity categories and for both radial and tangential source directions. As the pyramidal cells that give rise to the EEG potential are located in layer \uproman{5} of the gray matter \cite{Murakami2006}, eccentricities corresponding to 1-2mm distance to the skull are physiologically the most relevant. For eccentricities between 0.96 and 0.98 and both source directions CutFEM has average RDM/MAG values of 0.18\% and -0.06\%, comparable to UDGs 0.17\% and -0.2\% and significantly lower than \textit{Hex}'s 0.94\% and 1.57\%.
    
The most pronounced differences are at low eccentricities or when looking at magnitudes. CutFEM performance is similar for both radial and tangential source directions, UDG shows similar or slightly better results at low eccentricities. However, except for radial RDM's, UDG deteriorates faster at high eccentricities above 0.98. As both operate on the same cut mesh, the larger variance in the UDG results can most likely be explained by CutFEM's use of the ghost penalty stabilization. The overall largest absolute error values for CutFEM are 3.08 $\%$ RDM and 8.21 $\%$ MAG, underlining its performance with regard to outliers. Due to the similar numerical accuracy of CutFEM and UDG, we will only compare CutFEM and \textit{Hex} in the following scenarios.

\subsection{Sphere containing realistic brain}
The results in the previous section were achieved using analytically computed level sets. Deviating from this, we will now use use a semi-realistic case where realistic brain compartments are contained within spheres. 
Again, several different penalty parameters were tried, showing that a combination of  $\gamma = 40$ and a ghost penalty of $\gamma_g = 0.5$ yields good results for CutFEM.
	
The results are presented in Fig 3. Note that eccentricity is stated with respect to the distance to the skull. As source points are only inside the gray matter the number of source points at high eccentricities is much lower. The eccentricity groups 0.98, 0.985, 0.99, 0.995 were thus combined into one group containing 136 points.

	    \begin{figure*}[tp]
		\centering
		\subfloat{\includegraphics[width=1\linewidth]{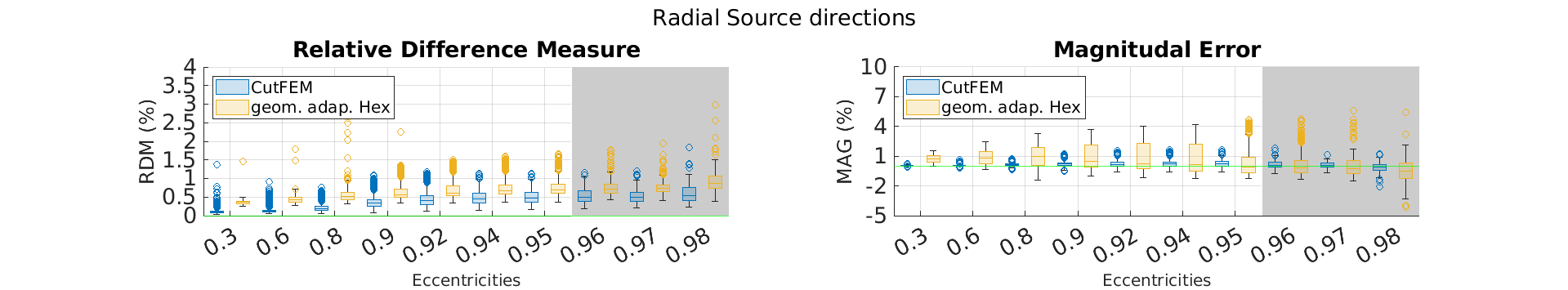}}
		\hfill
		\subfloat{\includegraphics[width=1\linewidth]{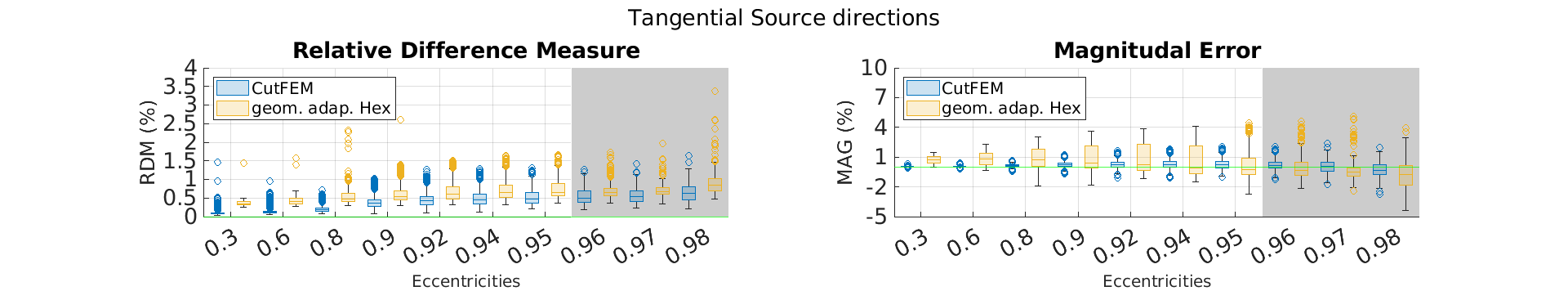}}
		\caption[EEG forward model errors for CG- and CutFEM - Realistic brain]{Overview of different EEG-errors for 5 layer continuous Galerkin- and CutFEM approaches using realistic brain compartments contained in spherical skull and scalp shells. Top: Errors for tangential source directions. Bottom: Errors for radial source directions. Errors are in percent and grouped by eccentricities. The green line marks optimal error values. The grey area indicates the physiologically most realistic eccentricities.}
	\end{figure*}
	
Much like before, CutFEM remains well below 1.5 and 2 percent RDM and MAG respectively, whereas \textit{Hex} has for nearly all eccentricities higher median values and more outliers going up to more than 1.5\% RDM and 4\% MAG. CutFEM is again more stable with regard to outliers and especially when looking at magnitudes, differences between the two methods are in the several percent range.
	
Overall, it can be stated that CutFEM is about as fast as and more accurate than \textit{Hex} and about as accurate as and faster than UDG.
	
\subsection{Realistic 5 compartment head model}
For the final part of this paper, two lead fields, one from CutFEM, one from hexahedral CG, were created using realistic 5 compartment head models including gray and white matter, CSF, skull and scalp tissues. Somatosensory evoked potentials were acquired from a medianus stimulation of the right hand.
	
\paragraph{Lead field differences} Before looking at inverse reconstructions, we will investigate the differences between the forward results. As the same source space and electrodes were used for both models, we can again compute MAG and RDM values. In the absence of an analytical solution, these measurements cannot capture errors but rather differences between the methods without making a clear statement which is the more accurate. 
	
For visualization purposes, for each gray matter centerpoint of the \textit{Hex} mesh the closest source point is identified, RDM and MAG are computed for each spatial direction and averages over the directions are calculated. The result can be seen in Fig. 4.  In both measures, the highest differences can be observed in inferior areas near the foramen magnum and optic channels or in superior areas. Overall the difference in potential distribution was 9.40 $\pm$ 4.15$\%$ and the difference in magnitude 18.94 $\pm$ 12.03$\%$. Interestingly, with a correlation coefficient of only 0.22,  high RDM values do not necessarily coincide with high MAG values.

	\paragraph{Reconstruction of somatosensory stimulation}
	Finally, the CutFEM lead field was used to perform a source reconstruction of the P20 component of an electric wrist stimulation. A dipole scan was conducted over the entire source space, the result of which can be seen in Fig. 4. $93.03\%$ of the data could be explained by a dipole with a strength of 5.8nAm. From the literature \cite{Buchner1994}, one expects the P20 component to be located in Brodmann Area 3b, that is in the anterior wall of the postcentral gyrus (and oriented towards the motor cortex).

	\section{Discussion}
	The purpose of this paper is to introduce CutFEM, an unfitted FEM for applications in EEG forward modeling. After discussing the mathematical theory behind CutFEM and implementational aspects, three progressively more realistic scenarios are introduced, ranging from a multi-layer sphere model to the reconstruction of somatosensory evoked potentials.     

	At similar computation times, CutFEM shows preferable results when compared to a geometry-adapted hexahedral CG-FEM \cite{Wolters2007} in both a shifted sphere scenario and a sphere model with realistic brain tissues. While CutFEM requires significantly less DOF, both methods require similar computation times due to the different number of solver iterations. Thus, a thorough investigation of different iterative solver techniques such as multigrid methods and possibly a modification of the ghost penalty will be part of future work.
 
    Compared to UDG \cite{Bastian2009}, it is shown that CutFEM combined with a ghost penalty leads to a decrease in outlier values at high eccentricities as well as a significant reduction in memory consumption and computation time. 
	
	Using a realistic five compartment head model, CutFEM correctly localizes the somatosensory P20 in the expected Brodmann area 3b. Especially in applications such as presurgical epilepsy diagnosis such accurate reconstructions might contribute significantly to the correct localization of the irritative zone \cite{Neugebauer2022}. The employed somatosensory experiment featured clear peaks and a high signal to noise ratio, making it an ideal candidate for an initial study. Further investigations and a larger study size are necessary to determine CutFEM's contribution to accurate source reconstructions when used with noisier data and/or more advanced inverse methods.

    In \cite{Vallaghe2010}, a trilinear immersed FEM approach was introduced that, like CutFEM, employs level sets as tissue surfaces. Rather than using a Nitsche-based coupling, continuity of the electric potential is enforced by modifying the trial function space. Compared to CutFEM, no free parameters such as $\gamma$ and $\gamma_G$ are introduced but the absence of overlapping submeshes means that there is no increased resolution in areas with complex geometries.

    In \cite{Windhoff2013,Nielsen2018} the process of building a tetrahedral mesh from segmentation data is investigated. Surface triangulations that are free of topological defects, self-intersections or degenerate angles have to be created before volumetric meshing. The authors show that it is possible to create such high quality surfaces and subsequent tetrahedral meshes for realistic head models, however they may come at the cost of modeling inaccuracies such as the separation of gray matter and skull by a thin layer of CSF.
    
    A main advantage of CutFEM is its flexibility with regard to the anatomical input data. Level sets can be created from a variety of sources, such as tissue probability maps, binary images or surface triangulations. This simplifies the question of how to create a mesh from segmentation data.
    
        \begin{figure*}[tp]
		\centering
		\subfloat[][]{\includegraphics[width=0.27\linewidth]{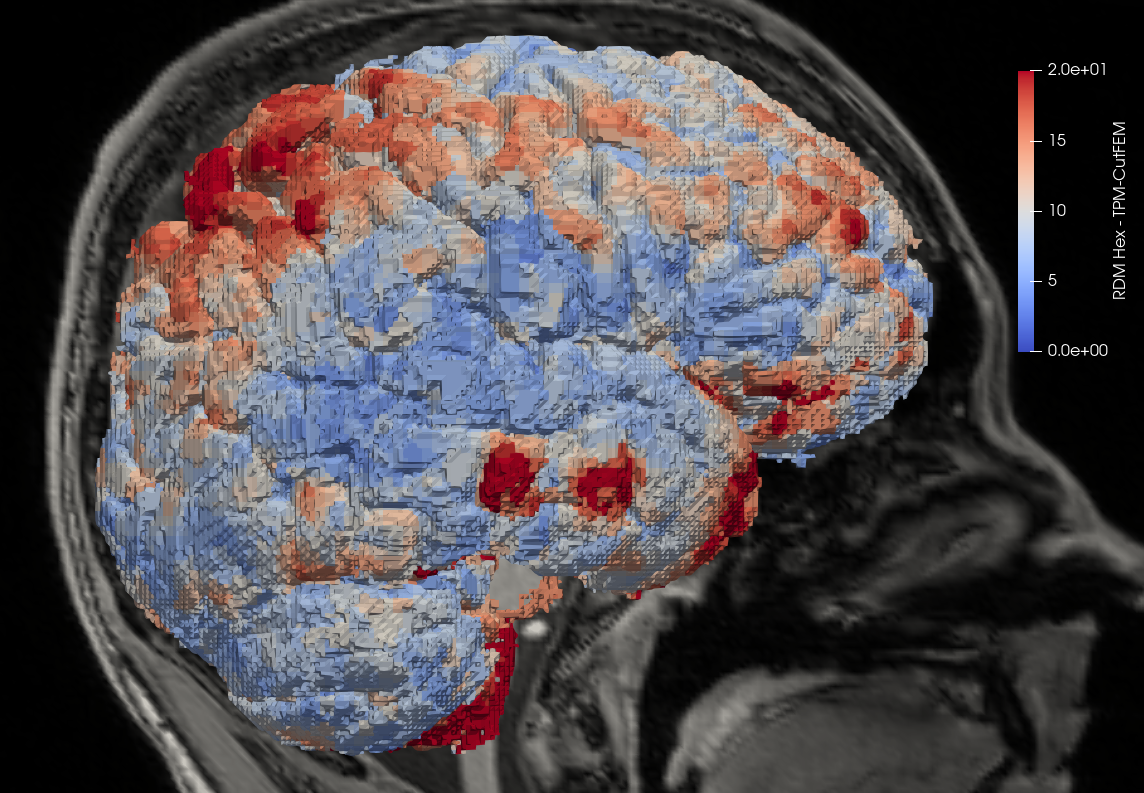}}
		\subfloat[][]{\includegraphics[width=0.27\linewidth]{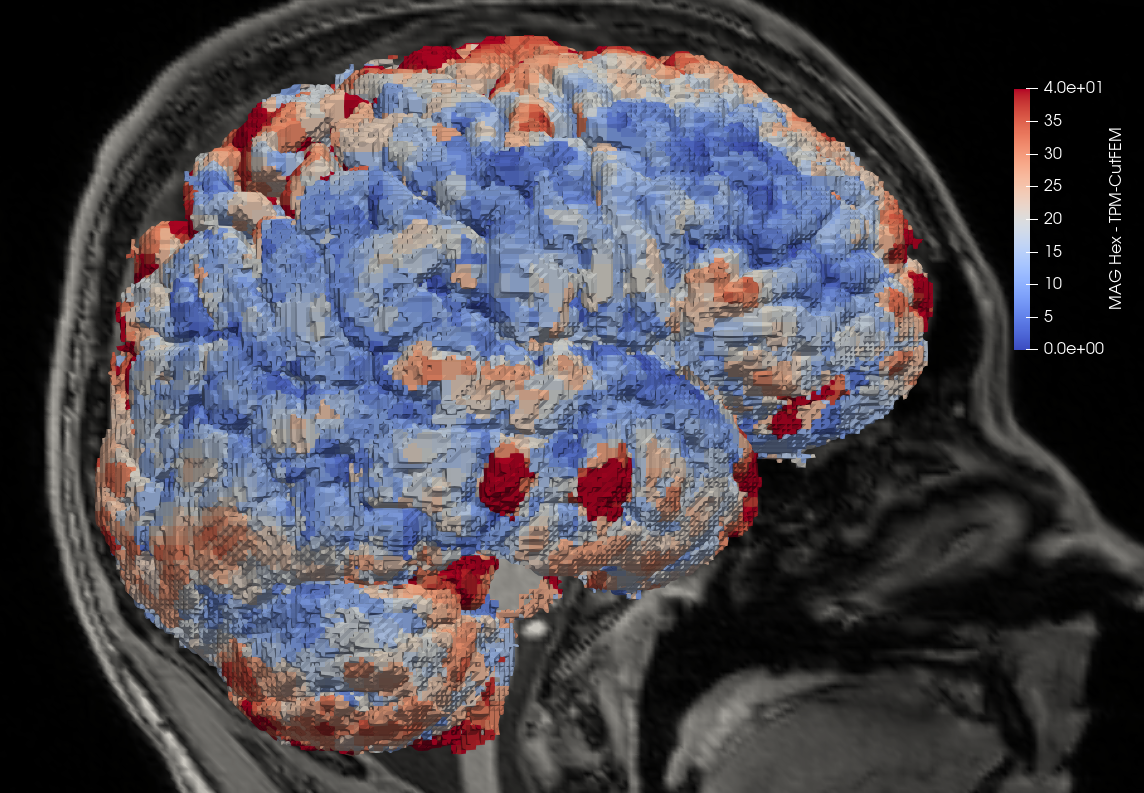}}
		\subfloat[][]{\includegraphics[width=0.27\linewidth]{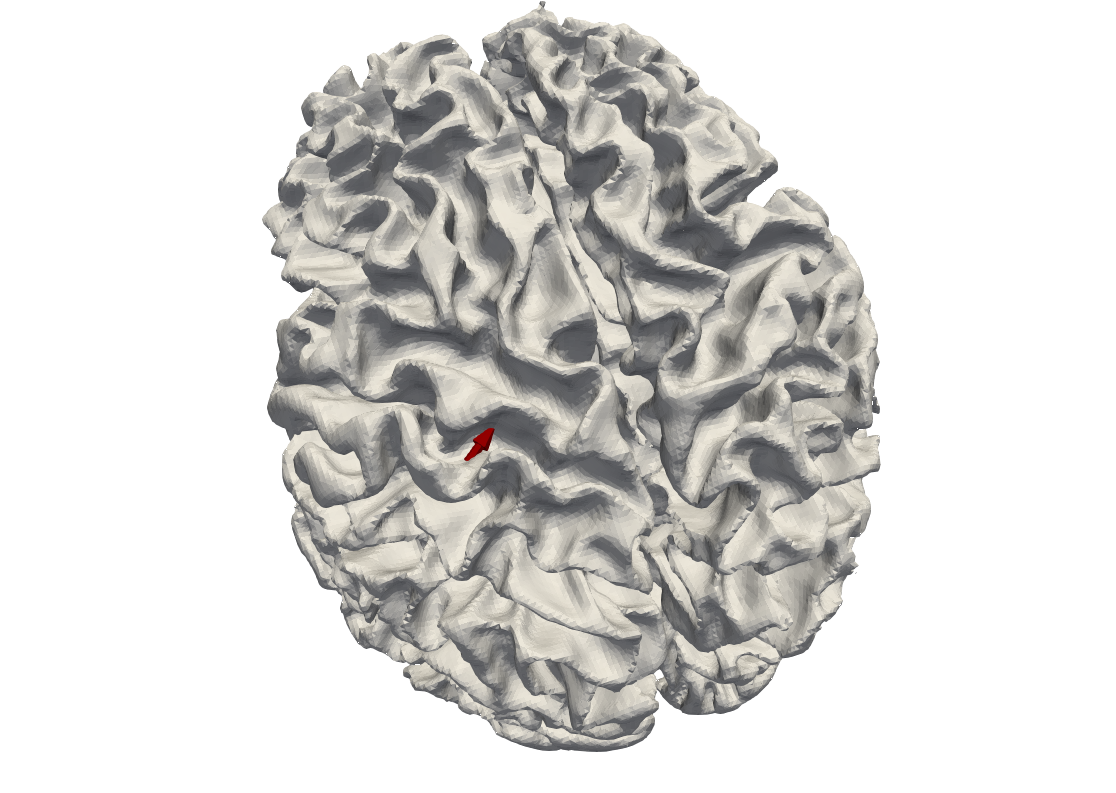}}
		\caption[RDM/MAG differences interpolated on gray matter + P20 reconstruction]{CutFEM vs \textit{Hex} lead field differences in distribution (a) and magnitude (b). Differences are interpolated onto gray matter. (c): The CutFEM based dipole reconstruction of the P20 medianus stimulation.}
	\end{figure*}
	
    \section{Conclusion}
	CutFEM performed well both when the underlying head model was created using analytical level sets or realistic segmentation results. Application to an inverse reconstruction of a somatosensory evoked potential yielded findings that are in line with the literature. The level sets underlying CutFEM impose few restrictions on the compartments, thus allowing for more simplified segmentation routines when compared to other FEM approaches using surface triangulations.

    \section{Author Contributions}
    \textbf{Conceptualization}: C. Engwer, A. Westhoff, Carsten H. Wolters.
    \textbf{Methodology}: C. Engwer, T. Erdbruegger,  A. Westhoff, C.H. Wolters.
    \textbf{Software}: C. Engwer, T. Erdbruegger, A. Westhoff.
    \textbf{Investigation}: T. Erdbruegger.
    \textbf{Writing – original draft}: T. Erdbruegger.
    \textbf{Writing – review and editing}: Y. Buschermoehle, C. Engwer, T. Erdbruegger,  J. Gross, M. Hoeltershinken, R. Lencer, J.O. Radecke, C.H. Wolters.
    \textbf{Supervision}: C. Engwer, C.H. Wolters.
    \textbf{Funding acquisition}: A. Buyx, J. Gross, R. Lencer, S. Pursiainen, F. Wallois, C.H. Wolters.
	\bibliographystyle{ieeetr}
	\bibliography{CutFEM_forward_modeling_for_EEG_source_analysis}{}
\end{document}